\documentclass[journal=jacsat,manuscript=article]{achemso}

\usepackage{chemformula} % Formula subscripts using \ch{}
\usepackage[T1]{fontenc} % Use modern font encodings
\usepackage{tabularx}
\usepackage{booktabs}
\usepackage{enumitem}

\author{Yajie Hao}
\affiliation{Institute of Fundamental and Frontier Sciences, University of Electronic Science and Technology of China, Chengdu 610051, China}

\author{Qiming Ding}
\email{dqiming94@pku.edu.cn}
\affiliation{Center on Frontiers of Computing Studies, Peking University, Beĳing 100871, China}%
\alsoaffiliation{School of Computer Science, Peking University, Beĳing 100871, China}%
\alsoaffiliation{ZQuantum (Beijing) Technology Co., Ltd, Beĳing 100871, China}

\author{Xiaoting Wang}
\email{xiaoting@uestc.edu.cn}
\affiliation{Institute of Fundamental and Frontier Sciences, University of Electronic Science and Technology of China, Chengdu 610051, China}

\author{Xiao Yuan}
\email{xiaoyuan@pku.edu.cn}
\affiliation{Center on Frontiers of Computing Studies, Peking University, Beĳing 100871, China}%
\alsoaffiliation{School of Computer Science, Peking University, Beĳing 100871, China}%

\title[An \textsf{achemso} demo]{Large-scale Efficient Molecule Geometry Optimization with Hybrid Quantum-Classical Computing}

\abbreviations{IR,NMR,UV}

\begin{document}

\begin{abstract}
% Accurately and efficiently predicting the equilibrium geometries of large molecules remains a central challenge in quantum computational chemistry, even with hybrid quantum--classical algorithms. Two major obstacles hinder progress: the large number of qubits required and the prohibitive cost of conventional nested optimization.
% In this work, a co-optimization framework based on Density Matrix Embedding Theory (DMET) is introduced to address these limitations. The proposed approach enables the determination of the equilibrium geometry of glycolic acid (C\(_2\)H\(_4\)O\(_3\))---a molecule of unprecedented complexity for quantum optimization---following validation on smaller benchmark systems such as H\(_4\) and H\(_2\)O\(_2\). This advancement represents a significant step toward extending quantum simulations beyond the small, proof-of-concept molecules that have dominated the field.
% The results demonstrate that the method achieves high accuracy while drastically reducing quantum resource requirements, establishing a practical route toward scalable quantum simulations. More broadly, this work paves the way for realizing quantum advantage in the \textit{in silico} design of complex catalysts and pharmaceuticals.

Accurately and efficiently predicting the equilibrium geometries of large molecules remains a central challenge in quantum computational chemistry, even with hybrid quantum--classical algorithms. Two major obstacles hinder progress: the large number of qubits required and the prohibitive cost of conventional nested optimization. 
In this work, we introduce a co-optimization framework that combines Density Matrix Embedding Theory (DMET) with Variational Quantum Eigensolver (VQE) to address these limitations. This approach substantially reduces the required quantum resources, enabling the treatment of molecular systems significantly larger than previously feasible. We first validate our framework on benchmark systems, such as H\(_4\) and H\(_2\)O\(_2\), before demonstrating its efficacy in determining the equilibrium geometry of glycolic acid  (C\(_2\)H\(_4\)O\(_3\))—a molecule of a size previously considered intractable for quantum geometry optimization. Our results show the method achieves high accuracy while drastically lowering computational cost. This work thus represents a significant step toward practical, scalable quantum simulations, moving beyond the small, proof-of-concept molecules that have historically dominated the field. More broadly, our framework establishes a tangible path toward leveraging quantum advantage for the in silico design of complex catalysts and pharmaceuticals.

\end{abstract}

\section{Introduction}

Molecular geometry lies at the heart of chemical physics: it dictates bond lengths and angles, governs molecular reactivity and stability, and shapes noncovalent interactions that underpin phenomena such as catalysis, crystal packing, and biomolecular recognition\cite{yang2021machine,delgado2021variational,chang2023integrating}. Precise geometry predictions are indispensable in diverse fields ranging from drug design and materials discovery to spectroscopy and reaction mechanism elucidation. However, for chemically relevant molecules containing tens or hundreds of atoms, the computational cost of traditional electronic structure methods—such as coupled-cluster or multireference configuration interaction—scales exponentially with system size\cite{szabo1996modern,levine2009quantum, kohn1999nobel}. This unfavorable scaling arises from the necessity to describe intricate many-electron correlations with high fidelity, an endeavor that rapidly becomes intractable as the number of electrons and basis functions grows\cite{mcardle2020quantum, nagy2024state}.

Quantum computing offers a fundamentally different computational paradigm, leveraging quantum superposition and entanglement to perform calculations beyond the reach of classical computers\cite{cao2019quantum,lanyon2010towards, mcardle2019digital,yuan2020quantum,aspuru2005simulated,xu2024mindspore,iijima2023towards}. Among various quantum algorithms, the Variational Quantum Eigensolver (VQE) has emerged as a promising candidate for approximating molecular ground-state energies on noisy intermediate-scale quantum (NISQ) devices\cite{cerezo2021variational,bauer2020quantum,lee2022variational,peruzzo2014variational,guo2024experimental,hempel2018quantum,lu2012quantum}. Nevertheless, despite remarkable demonstrations on small molecules, scaling VQE-based geometry optimization to larger, more chemically relevant systems faces two fundamental bottlenecks. First, the scarcity of available qubits limits the accessible size of quantum simulations, constraining the number of molecular orbitals that can be treated simultaneously\cite{nagy2024state,preskill2018quantum, cerezo2021variational,bharti2022noisy}. Second, the prohibitive computational cost of nested iterative optimization loops—in which the molecular geometry is updated only after a full quantum energy minimization—significantly slows convergence, especially when combined with the shot noise and finite sampling inherent to quantum devices. As a result, most quantum simulations to date have been confined to small systems of limited chemical complexity, leaving pharmaceutically and industrially relevant molecules largely unexplored\cite{kandala2017hardware,nam2020ground,boyn2021quantum,google2020hartree,hu2022benchmarking,cao2023ab,ding2024molecular, cao2021towards}.

To address these challenges, we introduce a novel hybrid quantum-classical co-optimization framework built upon Density Matrix Embedding Theory (DMET)\cite{wouters2016practical,knizia2013density}. DMET enables the systematic partitioning of a large molecular system into smaller, computationally tractable fragments while rigorously preserving the entanglement and electronic correlations between them. This fragmentation dramatically reduces the number of qubits required for the quantum simulation without sacrificing accuracy\cite{shang2023towards,li2022toward}. In our framework, DMET is tightly integrated with VQE in a direct co-optimization procedure, where both the molecular geometry and the quantum variational parameters are optimized simultaneously. This design eliminates the expensive outer optimization loop over geometries, thereby accelerating convergence and reducing the number of quantum evaluations required. By combining DMET’s scalability with VQE’s capability for accurate correlated energy estimation, our approach overcomes the dual barriers of qubit limitations and iterative cost.

We validate our method on benchmark molecules such as H\(_4\) and H\(_2\)O\(_2\), demonstrating that the co-optimization framework achieves high-fidelity equilibrium geometries with substantially fewer quantum resources compared to conventional approaches. We then extend the methodology to glycolic acid  (C\(_2\)H\(_4\)O\(_3\)), a chemically and biologically significant molecule whose complexity places it well beyond the reach of prior quantum algorithms. To the best of our knowledge, this represents the first successful quantum algorithm-based geometry optimization of a molecule of this scale and complexity. Our results not only match the accuracy of classical reference methods but also drastically reduce quantum resource demands. This breakthrough marks an important step toward making realistic, large-scale molecular geometry optimization feasible on near-term quantum devices, opening a pathway toward the quantum simulation of complex chemical, biological, and materials systems previously considered intractable.

\section{Preliminaries}

Under the Born-Oppenheimer approximation\cite{mcardle2020quantum}, the general Hamiltonian of a quantum chemical system can be expressed as:
\begin{equation}
\hat{H} = E_{\text{nuc}} + \sum_{pq} \hat{D}_{pq} + \sum_{pqrs} \hat{V}_{pqrs},
\end{equation}
where \( E_{\text{nuc}} \) is the scalar nuclear repulsion energy, \( \hat{D}_{pq} = d_{pq} \hat{a}_p^\dagger \hat{a}_q \) and \( \hat{V}_{pqrs} = \frac{1}{2} h_{pqrs} \hat{a}_p^\dagger \hat{a}_q^\dagger \hat{a}_r \hat{a}_s \) are the one-body and two-body interaction operators, respectively, \( \hat{a}_p \) (\( \hat{a}_p^\dagger \)) is the fermionic annihilation (creation) operator for the \( p \)th orbital, and \( \{d_{pq}\} \) and \( \{h_{pqrs}\} \) are the corresponding one- and two-electron integrals computed classically. Here, the molecular spin-orbitals are denoted as \( p, q, r, s \).

For large-scale simulations, the number of qubits and the circuit depth become unaffordable on near-term quantum devices. In order to reduce the usage of the computational resources, it is necessary to resort to the DMET strategy\cite{knizia2012density}. In the following, we will introduce the standard DMET approaches\cite{shang2023towards,cao2023ab,li2022toward,kawashima2021optimizing}.

Consider a molecular system divided into fragment \( A \) with basis \( \{|\psi_i^A\rangle\} \) of dimension $d_A$ and environment \( B \) with basis \( \{|\psi_j^B\rangle\} \) of dimension $d_B$. As shown in Fig.~\ref{fig:vqe-dmet-geo_loop}, when an atom is selected as a fragment \( A \), the remaining part is regarded as the environment \( B \). The full quantum state in the \( \{|\psi_i^A\rangle |\psi_j^B\rangle\} \) basis can be represented as:
\begin{equation}
|\Psi\rangle = \sum_{i,j} \Psi_{i,j} |\psi_i^A\rangle |\psi_j^B\rangle.
\end{equation}

This representation can be greatly simplified by considering the entanglement between the two parts. Specifically, the quantum state \( |\Psi\rangle \) can be decomposed into a rotated basis \( \{|\tilde{\psi}_a^A\rangle |\tilde{\psi}_a^B\rangle\} \), corresponding to the Schmidt decomposition of bipartite states:
\begin{equation}
|\Psi\rangle = \sum_{a=1}^{d_k} \lambda_a |\tilde{\psi}_a^A\rangle |\tilde{\psi}_a^B\rangle,
\end{equation}
where the $d_k=\min(d_A,d_B)$. Next, the Hamiltonian of fragment $A$
embedded in bath $B$ can be defined by projecting the full Hamiltonian \( \hat{H} \) into the space spanned by the basis of the fragment and bath:
\begin{equation}
\hat{H}_{\text{emb}} = \hat{P} \hat{H} \hat{P}.
\end{equation}
Here, the projector \( \hat{P} \) is defined as:
\begin{equation}
\hat{P} = \sum_{ab} |\tilde{\psi}_a^A \tilde{\psi}_b^B\rangle \langle \tilde{\psi}_a^A \tilde{\psi}_b^B |.
\end{equation}
We note that the embedded Hamiltonian can be represented by the one-electron integral and the two-electron integral as follows
\begin{equation}
\begin{split}
\label{eq:fragment_ham}
\hat{H}_{\text{emb}} = & \sum^{L_A+L_B}_{ps}[d_{ps} + \sum^L_{qr}(h_{psrq}-h_{prqs})D^{env}_{qr}]\hat{a}_p^\dagger \hat{a}_s +\sum^{L_A+L_B}_{pqrs} h_{pqrs} \hat{a}_p^\dagger \hat{a}_q^\dagger \hat{a}_r \hat{a}_s,
\end{split}
\end{equation}
where $L_A$ is the number of orbitals in the fragment, $L_B$ is the number of bath orbitals, $L$ is the number of orbitals in the entire molecule, and the environment density matrix of fragment $A$ is $D^{env}_{qr} = \sum_{p\in env}C_{qp}C^\dagger_{pr}$ with $C$ being the molecular orbital coefficients obtained from the mean-field
calculation of the entire molecule.

We can find that if \( |\Psi\rangle \) is the ground state of a Hamiltonian \( \hat{H} \), it must also be the ground state of \( \hat{H}_{\text{emb}} \). This indicates that the solution of a small embedded system is equivalent to that of the full system. This decomposition reduces the computational problem of the entire system into a more tractable, smaller-scale problem. 

However, the exact wavefunction of the molecular system is typically unavailable in advance, necessitating the use of approximations. A natural choice for such an approximation is the wavefunction derived from a low-level mean-field theory, such as Hartree-Fock (HF) calculations. This approximate wavefunction serves as a foundation to construct the environmental bath, enabling the application of a high-level theory to solve the reduced problem. So, in order to get a high-level calculation for fully molecules, we need to optimize the embedding of a bath. 

In DMET, a high-level calculation for each fragment is carried out individually until self-consistency has been attained according to a certain criterion: The number of electrons in the fragments sum up to the total number of electrons of the full system. A global potential $\mu$ is applied to fix the embedded Hamiltonian:
\begin{equation}
\begin{split}
H_{emb}\xleftarrow{}H_{emb} - \mu \sum_{p \in A}\hat{a}_p^\dagger \hat{a}_p.
\end{split}
\end{equation}

Once the embedding Hamiltonian $H_{emb}$ ground state $\psi_A$ has been obtained, the 1- and 2-RDMs of fragement $A$ are defined as
\begin{equation}
\begin{split}
D^A_{pq} &=\langle\psi_A|\hat{a}_p^\dagger \hat{a}_q|\psi_A\rangle,\\
P^A_{pqrs} &=\langle\psi_A|\hat{a}_p^\dagger \hat{a}_q^\dagger \hat{a}_r \hat{a}_s|\psi_A\rangle.
\end{split}
\end{equation}
Next, we can get the DMET cost function is written as
\begin{equation}
\begin{split}
\label{eq:dmet_cost}
L(\mu) = \left(\sum_{A}\sum^{L_A}_{r} D^A_{rr}(\mu)+N_{mf}-N_{occ}\right)^2,
\end{split}
\end{equation}
where $N_{mf}$ is the number of electrons in the inactive orbitals obtained at the mean-field level and $N_{occ}$ is the total number of electrons.
It can be employed to optimize the global potential $\mu$ until the DMET cost function converges\cite{wouters2016practical}.

The total DMET energy is calculated by summing the fragment energy for each
fragment, which is obtained according to the equation
\begin{equation}
\begin{split}
E_{tot}=\sum_A E^A+E_{nuc},
\end{split}
\end{equation}
where the fragment energy $E_A$ as
\begin{equation}
\begin{split}
\label{eq:fragment_energy}
E^A = \sum^{L_A}_p \bigg( \sum^{L_A+L_B}_{s}\big[d_{ps} + \frac{1}{2}\sum^L_{qr}(h_{psrq}-h_{prqs})D^{env}_{qr}\big]D_{ps} + \frac{1}{2}\sum^{L_A+L_B}_{qrs} h_{pqrs}P_{pqrs} \bigg).
\end{split}
\end{equation}

To obtain the solution of the high-level embedding Hamiltonian, we use Variational Quantum Eigensolvers (VQEs)\cite{peruzzo2014variational,hu2022benchmarking,bharti2022noisy,kandala2017hardware}. The key component in VQE is to design an appropriate circuit ansatz to approximate the unknown ground state of the chemical system. Here, we use the Unitary Coupled-Cluster Singles and Doubles (UCCSD) ansatz\cite{barkoutsos2018quantum,hu2022benchmarking,grimsley2019trotterized}, which effectively considers excitations and de-excitations above a reference state:
\begin{equation}
|\psi\rangle = e^{T - T^\dagger} |\psi_0\rangle,
\end{equation}
where the reference state \( |\psi_0\rangle \) is chosen as the Hartree-Fock ground state in the basis of the embedded system, and the cluster operator \( T \) is truncated at single and double excitations, has the form
\begin{equation}
T(\vec{\theta}_A)=\sum_{pq}\theta_{pq}T_{pq}+\sum_{pqrs}\theta_{pqrs}T_{pqrs},
\end{equation}
where the one-body and two-body term are defined as $T_{pq}=\hat{a}_p^\dagger \hat{a}_q$ and $T_{pqrs}=\hat{a}_p^\dagger \hat{a}_q^\dagger \hat{a}_r\hat{a}_s$.
Next, we minimize the energy of the embedded system for each fragment $A$:
\begin{equation}
E^A_{\text{emb}} = \min_{\vec{\theta}_A} \langle \psi(\vec{\theta}_A) | \hat{H}^A_{\text{emb}} | \psi(\vec{\theta}_A) \rangle.
\end{equation}
It should be mentioned that the optimization of $\vec{\theta}_A$ is often performed through gradient descent, where the gradient of $\vec{\theta}_A$ can be obtained through parameter shift rules\cite{mitarai2018quantum,schuld2019evaluating}. Through such a gradient descent, until the VQE cost function converges, we can obtain the high-level wavefunction and thus get the reduced density matrices of the simplified system.

Finally, after the VQE cost function and DMET cost function converge, the energy $E_A$ corresponding to each fragment is calculated, and then $E_{tot}=\sum_A E^A+E_{nuc}$ is the ground state energy of the entire molecule.

\section{Methods}

Now, we introduce our method for optimizing molecular geometries based on a hybrid VQE-DMET method.
The optimization of molecular geometries through VQE involves defining a parameterized Hamiltonian that characterizes the electronic structure of a molecule. This Hamiltonian, expressed in the second quantization framework for a set of nuclear coordinates \( x \), is represented as follows:

\begin{equation}
\label{eq:x_0_ham}
\hat{H}(\vec{x}) = E_{\text{nuc}}(\vec{x}) + \sum_{pq} d_{pq}(\vec{x}) \hat{a}_p^\dagger \hat{a}_q + \sum_{pqrs} \frac{1}{2} h_{pqrs}(\vec{x}) \hat{a}_p^\dagger \hat{a}_q^\dagger \hat{a}_r \hat{a}_s. 
\end{equation}
Here, the terms \( d_{pq}(\vec{x}) \) and \( h_{pqrs}(\vec{x}) \) represent the one- and two-electron Coulomb integrals, computed from the geometry determined by the parameters $\vec{x}$.

\begin{figure}
% \centering
\includegraphics[width=0.9\textwidth]{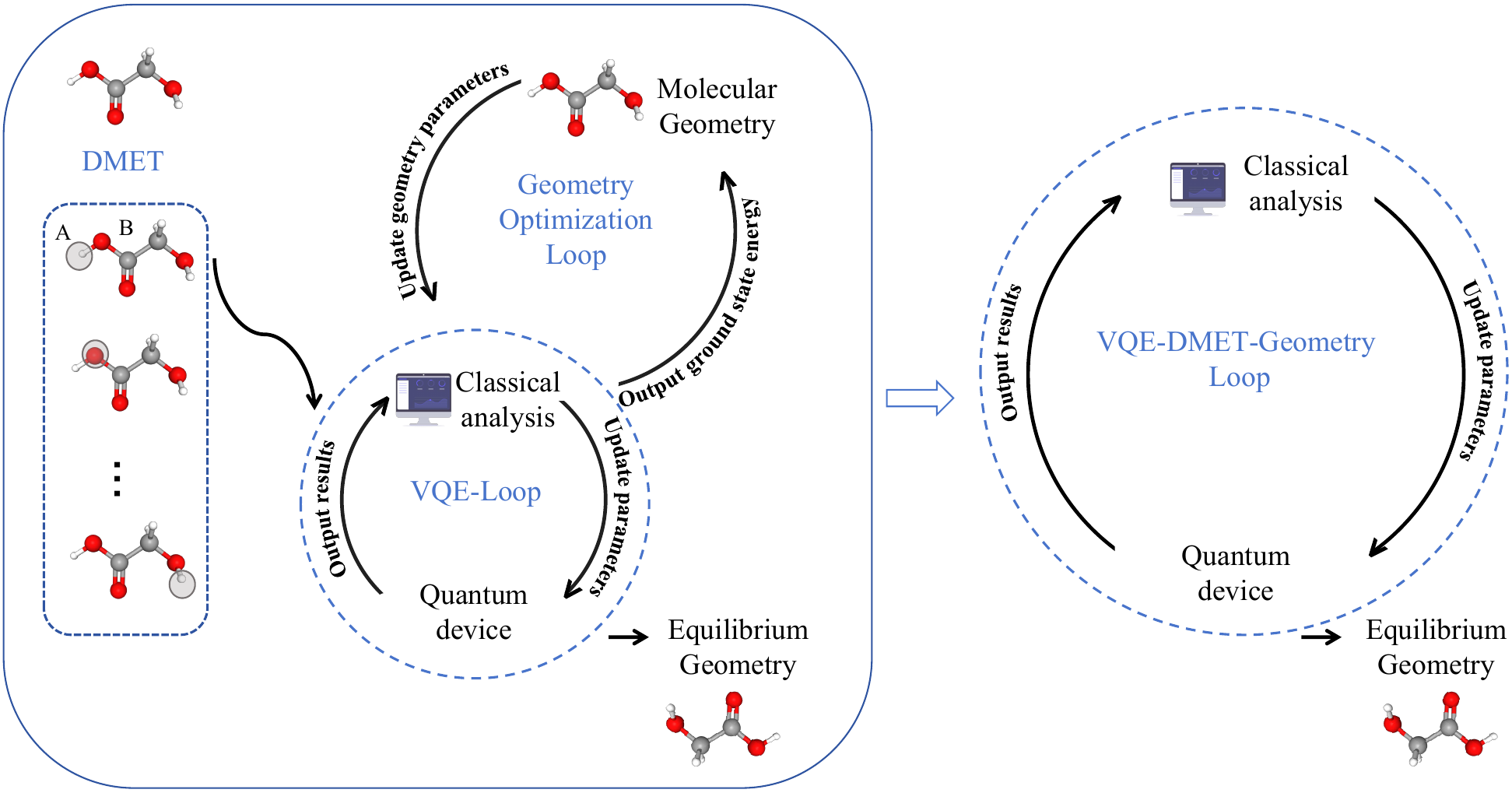}
\caption{Comparison of two VQE–DMET-based geometry optimization strategies. Left: the conventional nested scheme\cite{cao2021towards}, where an outer geometry optimization loop updates molecular geometry based on energies obtained from an inner VQE–DMET calculation, incurring high computational cost. Right: the proposed direct co-optimization approach, which updates both geometry and VQE variational parameters simultaneously, enabling more efficient convergence to equilibrium.}
\label{fig:vqe-dmet-geo_loop}
\end{figure}

The standard approach for optimizing molecular geometry, depicted on the left side of Fig.~\ref{fig:vqe-dmet-geo_loop}, involves a two-level iterative process. An outer geometry optimization loop adjusts the molecular parameters $\vec{x}$, relying on the ground state energy provided by an inner VQE-DMET loop. Within the VQE-DMET loop, a classical optimizer interacts with a quantum device to find the ground state energy for the current molecular configuration. This energy then guides the geometry optimization, iteratively moving towards equilibrium. However, this nested iterative scheme demands significant computational resources due to the repeated execution of the inner VQE-DMET loop for each geometry update.

To address this, we propose an alternative framework, shown on the right side of Fig.~\ref{fig:vqe-dmet-geo_loop}. Our VQE-DMET-Geometry loop directly optimizes both the geometry parameters $\vec{x}$ and the VQE variational parameters simultaneously, aiming for a more efficient convergence to the equilibrium geometry and ground state energy.

In this framework, let \( |\psi(\beta)\rangle \) represent the trial state of the qubit system, parameterized by a set of circuit parameters \( \beta \). The VQE cost function, defined as the expectation value of the Hamiltonian in this trial state, is given by:

\begin{equation}
g(\vec{\beta}, \vec{x}) = \langle \psi(\vec{\beta}) | H(\vec{x}) | \psi(\vec{\beta}) \rangle.
\end{equation}
The central objective is to minimize this cost function with respect to both \( \vec{\beta} \) and \( \vec{x} \):
\begin{equation}
E = \min_{\vec{\beta}, \vec{x}} g(\vec{\beta}, \vec{x}).
\label{eq:vqe_x_beta}
\end{equation}
Through this process,  the optimal geometric structure of the molecule and its corresponding ground state energy are obtained.

It should be noted that Eq.~(\ref{eq:vqe_x_beta}) does not include the DMET framework. When DMET is integrated into the optimization, the corresponding objective function is modified as follows:
\begin{equation}
\begin{split}
\label{eq: dmet_geo}
    \min_{\vec{\theta}, \vec{x}} g(\vec{\theta}, \vec{x}) =& \min_{\vec{\theta}, \vec{x}}\langle \psi(\vec{\theta}) | H(\vec{x}) | \psi(\vec{\theta}) \rangle,\\
    =&\min_{\vec{\theta}_A,\vec{x}} (\sum_A E^A(\vec{\theta}_A,\vec{x})  + E_{unc}(\vec{x})).
\end{split}
\end{equation}
Here $\vec{\theta}$ contains the different variational parameters corresponding to all fragments. For simplicity of expression, we have omitted the outer DMET loop optimization of the global potential $\mu$, but this remains an indispensable part of the overall algorithm. Furthermore, the optimization of $\vec{\theta}$ must satisfy the ground state constraints of the embedding Hamiltonian. Consequently, we arrive at the following constrained optimization problem
\begin{equation}
\begin{split}
    &\min_{\vec{\theta}_A,\vec{x}}  \left(\sum_A E^A(\vec{\theta}_A,\vec{x})  + E_{unc}(\vec{x})\right),\\
    \text{s.t.}  \quad &\vec{\theta}_A = \arg\min_{\vec{\theta'}_A} \left(\sum_A \langle \psi(\vec{\theta'}_A) | \hat{H}^A_{\text{emb}}(\vec{x}) | \psi(\vec{\theta'}_A) \rangle \right).
\end{split}
\end{equation}

To solve such a constrained optimization problem, we employ an alternating optimization strategy. The optimization process iteratively updates the parameters $\vec{x}$ and $\vec{\theta}$ by solving the following subproblems:
\begin{align}
\vec{x}^{(k+1)} &= \arg\min_{\vec{x}} \left( \sum_A E^A(\vec{\theta}^{(k)}_A, \vec{x})  + E_{\text{unc}}(\vec{x}) \right), \label{eq:opt_E_A_revised} \\
\vec{\theta}^{(k+1)} &= \arg\min_{\vec{\theta}} \left( \sum_A \langle \psi(\vec{\theta}_A) | \hat{H}^A_{\text{emb}}(\vec{x}^{(k+1)}) | \psi(\vec{\theta}_A) \rangle \right). \label{eq:opt_H_emb_revised}
\end{align}
Specifically, at the $k$-th iteration, we first fix $\vec{\theta}^{(k)}$ and solve the optimization problem in Eq.~\eqref{eq:opt_E_A_revised} to obtain the updated $\vec{x}^{(k+1)}$. Subsequently, with $\vec{x}^{(k+1)}$ held constant, we solve Eq.~\eqref{eq:opt_H_emb_revised} to find the updated $\vec{\theta}^{(k+1)}$. This two-step process is repeated until a convergence criterion is met.

Here we use gradient-based optimization to optimize $\vec{\theta}$ and $\vec{x}$, which is expressed as follows
\begin{equation}
\begin{split}
\label{eq: grad}
\vec{\theta}_A & =\vec{\theta}_A - \eta \frac{\partial \langle \psi(\vec{\theta}_A) | \hat{H}^A_{\text{emb}}(x) | \psi(\vec{\theta}_A) \rangle}{\partial \vec{\theta}_A},\\
\vec{x} &= \vec{x} - \eta \frac{\partial \left(\sum_A E^A(\vec{\theta}_A,\vec{x}) + E_{nuc}(\vec{x})\right) }{\partial \vec{x}},
\end{split}
\end{equation}
where $\eta$ is the learning rate, which controls the step size of the optimization.

\begin{figure}
% \centering
\includegraphics[width=8cm]{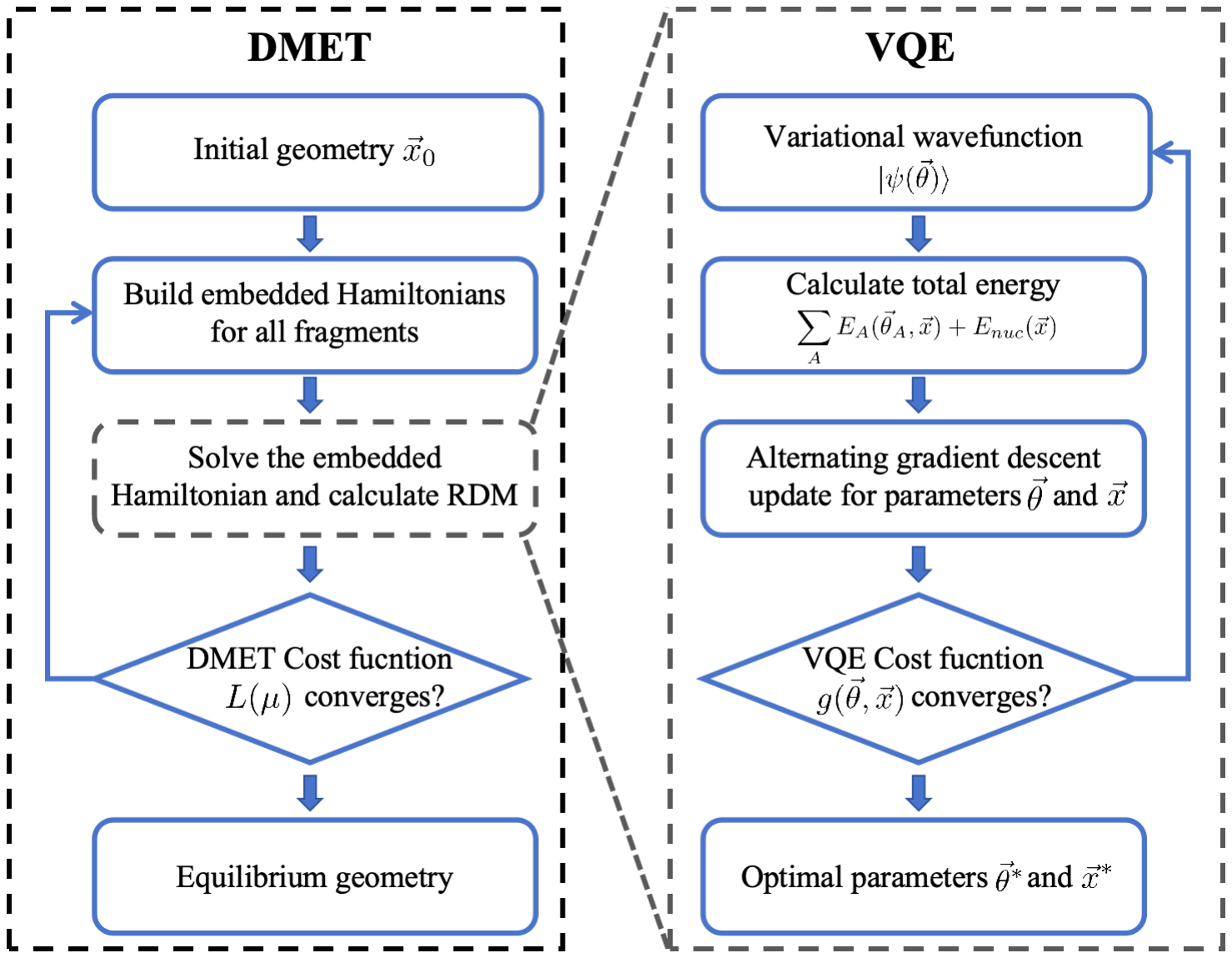}
\caption{The workflow for the VQE-DMET-Geometry Co-optimization Algorithm.  The left (DMET) loop involves building and solving embedded Hamiltonians to calculate the RDM, converging based on $L(\mu)$. Concurrently, the right (VQE) loop constructs a variational wavefunction, calculates total energy, and updates VQE parameters $\vec{\theta}$ and geometry $\vec{x}$ via alternating gradient descent, converging based on $g(\vec{\theta}, \vec{x})$. The process yields the equilibrium geometry and optimal parameters $\vec{\theta}^*$ and $\vec{x}^*$.}
\label{fig:alog_flow}
\end{figure}

The next crucial aspect is obtaining the gradient information required for the parameter updates in Eq.~(\ref{eq: grad}). The derivative $\partial \langle \psi(\vec{\theta}_A) | \hat{H}^A_{\text{emb}}(\vec{x}) | \psi(\vec{\theta}_A) \rangle / \partial \vec{\theta}_A$ can be calculated using parameter-shift rules\cite{mitarai2018quantum,schuld2019evaluating}. And the term $\partial \left(\sum_A E^A(\vec{\theta}_A,\vec{x}) + E_{nuc}(\vec{x})\right) / \partial x$ can be calculated by the following expression:
\begin{equation}
\begin{split}
\label{eq: grad_x_cal}
\frac{\partial \left(\sum_A E^A(\vec{\theta}_A,\vec{x}) + E_{nuc}(\vec{x})\right) }{\partial \vec{x}}=\sum_A \frac{\partial E^A(\vec{\theta}_A,\vec{x})}{\partial \vec{x}} +\frac{\partial E_{nuc}(\vec{x})}{\partial \vec{x}}.
\end{split}
\end{equation}
According to the expression Eq.~(\ref{eq:fragment_energy}) of $E_A$ and Hellmann–Feynman theorem\cite{stanton1962hellmann, politzer2018hellmann}, it can be found that Eq.~(\ref{eq: grad_x_cal}) is simplified to calculate $\frac{\partial h_{pq}(\vec{x})}{\partial \vec{x}}$, $\frac{\partial h_{pqrs}(\vec{x})}{\partial \vec{x}}$ and $\frac{\partial E_{nuc}(\vec{x})}{\partial \vec{x}}$. These derivatives can be entirely computed via classical differential calculations, thereby avoiding additional quantum computing resource overhead.

Thus far, we can get a complete algorithm flow combining DMET and VQE to optimize molecular geometry as follows (see also Fig.~\ref{fig:alog_flow}).
\begin{enumerate}[label={Step \arabic*.}]
    \item Set the initial geometry of the molecule and characterize it by the nuclear coordinates $\vec{x}_0$, and obtain the Hamiltonian as Eq.~(\ref{eq:x_0_ham}).
    
    \item Set the initial global potential $\mu$, then divide the molecular system into fragments and obtain the embedded Hamiltonian of each fragment as Eq.~(\ref{eq:fragment_ham}). Note that this embedded Hamiltonian includes the geometry parameter $\vec{x}_0$ and global potential $\mu$.
    
    \item Variational quantum circuit $\psi(\vec{\theta})$ on the quantum device for each fragment to get $\langle \psi(\vec{\theta}_{\text{frag}}) | \hat{H}^{\text{frag}}_{\text{emb}}(\vec{x}) | \psi(\vec{\theta}_{\text{frag}}) \rangle$ and $E_{\text{frag}}(\vec{\theta}_{\text{frag}},\vec{x_0})$.
    
    \item Update the corresponding parameters $\vec{\theta}$ and $\vec{x}$ according to Eq.~(\ref{eq: grad}), and then return to the Step 3 until the cost function Eq.~(\ref{eq: dmet_geo}) converges.
    
    \item Check whether the cost function Eq.~(\ref{eq:dmet_cost}) of DMET converges. If so, get the optimal geometry parameter $\vec{x}^*$. Otherwise, adjust the global potential and return to Step 2.
\end{enumerate}

This entire algorithmic process incorporates two nested loops for parameter updates. The inner loop, driven by VQE, optimizes both $\vec{\theta}$ and  
$\vec{x}$ simultaneously. The outer loop, governed by DMET, optimizes the global potential $\mu$. This framework enables large-scale molecular geometry optimization to be performed effectively on small-scale quantum devices.

\section{Results and discussion}

Now, we introduce the application of the proposed quantum-classical algorithm to determine the ground-state equilibrium geometries of H\(_4\), H\(_2\)O\(_2\), and glycolic acid C\(_2\)H\(_4\)O\(_3\) molecules. We first validate our approach on the well-characterized small molecular systems, H\(_4\) and H\(_2\)O\(_2\), before extending its application to the more complex glycolic acid C\(_2\)H\(_4\)O\(_3\), a relevant $\alpha$-hydroxy acid.

\begin{figure}
% \centering
\includegraphics[width=8cm]{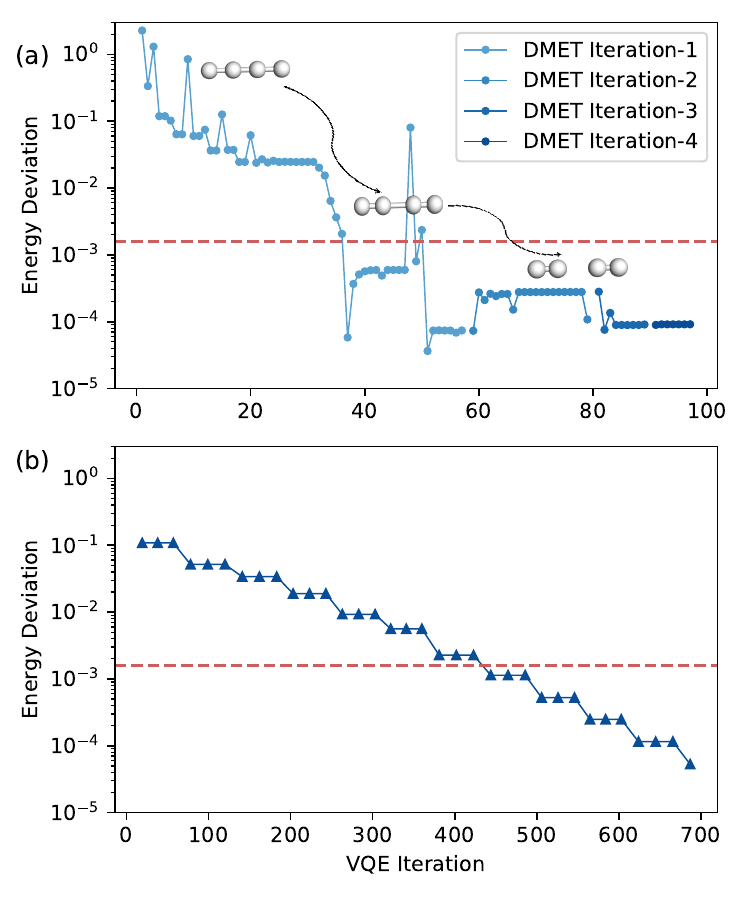}
\caption{(a) Convergence of our proposed algorithm for H\(_4\) optimization, showing energy deviation versus VQE iterations and the dissociation from a linear chain to two H\(_2\) molecules. (b) Optimization iterations of the standard nested loop method (conceptually depicted in Fig.~\ref{fig:vqe-dmet-geo_loop}). Our framework achieves convergence with significantly fewer iterations and computational resources compared to the standard method.}
\label{fig:H4}
\end{figure}

For the H\(_4\) molecule, we employed the STO-3G minimal basis set. Under the Jordan–Wigner mapping, a full simulation of this system typically requires 8 qubits. While applying the DMET strategy to such a small system may not offer substantial practical advantages in terms of computational resource savings, its inclusion here serves to effectively demonstrate the methodology. Within the DMET framework, each atom was treated as a separate fragment, resulting in a total of 4 fragments. Each fragment calculation required a maximum of 4 qubits, thereby effectively reducing the qubit requirement for the overall H\(_4\) simulation from 8 qubits to 4 qubits within the embedded approach.

\begin{figure*}
\includegraphics[width=\textwidth]{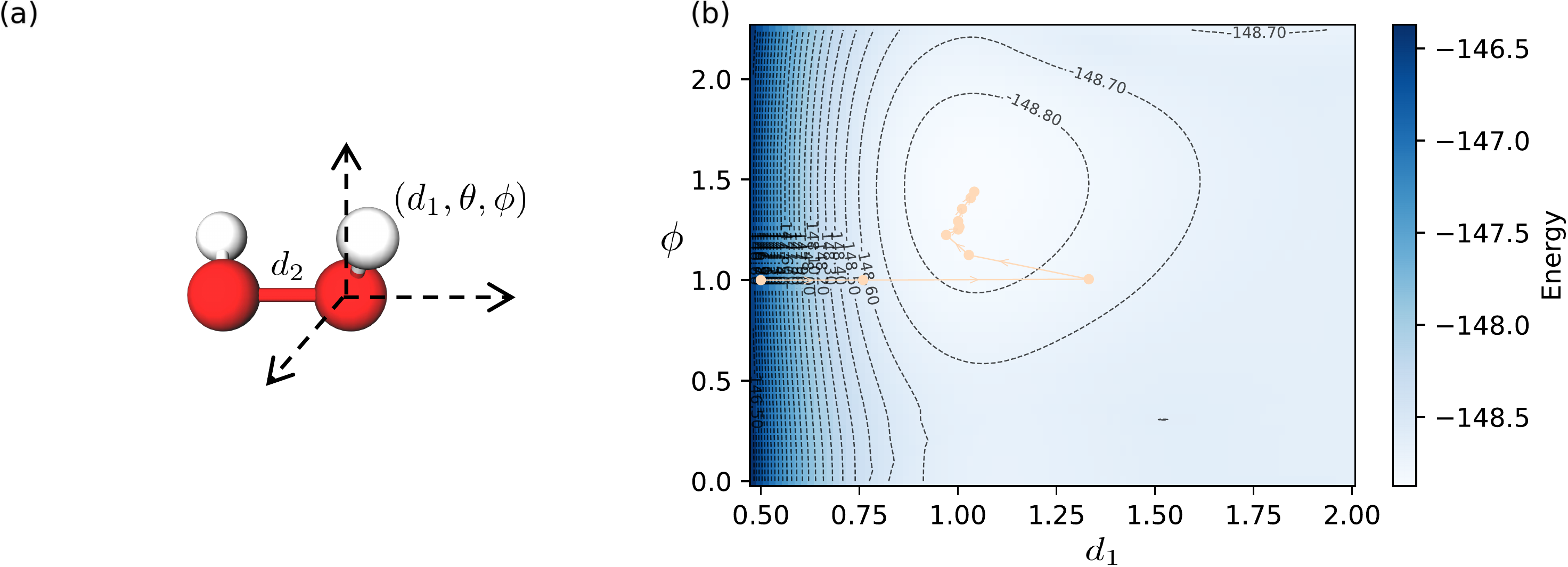}
\caption{Exploration of molecular geometries and configurations using our algorithm. (a) Depiction of the H\(_2\)O\(_2\) molecule with its key geometric parameters ($d_1$, $d_2$, $\theta$, $\phi$). (b) Trajectory of the H\(_2\)O\(_2\) optimization in the $d_1$-$\phi$ parameter space, showing convergence to a low-energy region (heatmap is a classical reference). }
\label{fig:H2O2}
\end{figure*}

The molecular geometry optimization for H\(_4\) commenced from an initial linear configuration of four hydrogen atoms. The three bond lengths within this linear chain were chosen as the optimization parameters, all initially set to 1.0 Å. The convergence process of our proposed algorithm framework is depicted in Fig.~\ref{fig:H4}(a). Here, the horizontal axis represents the number of VQE iterations, while the vertical axis shows the energy deviation relative to the H\(_4\) molecule's equilibrium energy. Fig.~\ref{fig:H4}(a) clearly illustrates the evolution of the H\(_4\) geometric configuration: starting from an initial linear arrangement with equal bond lengths, the optimization trajectory progressively leads to a configuration resembling two dissociated H\(_2\) molecules. This observation demonstrates the spontaneous dissociation of the H\(_4\) molecule during the geometry optimization process. For comparison, Fig.~\ref{fig:H4}(b) presents the optimization iteration process using a standard nested loop method, as conceptually depicted on the left side of Fig.~\ref{fig:vqe-dmet-geo_loop}. A direct comparison with our framework reveals that the standard nested loop method demands significantly more iterations and consumes substantially greater computational resources to achieve convergence, highlighting the efficiency of our approach.

\begin{table*}
    \centering
    \caption{Comparison of initial qubit counts with DMET-reduced qubit counts for H\(_4\), H\(_2\)O\(_2\), and glycolic acid C\(_2\)H\(_4\)O\(_3\) molecules. The table also presents the optimized geometric parameters obtained from Our Work alongside their respective reference values and the calculated deviation.}
    \label{tab:optimized_geometries}
    \begin{tabularx}{\textwidth}{
            X % Molecule 
            X % Initial Qubits
            X % DMET Qubits
            X % Parameter
            X % Our Work 
            X % Reference 
            X % Deviation 
        }
        \toprule
        Molecule &Initial Qubits & DMET Qubits & Parameter & {Our Work} & {Reference} & {Deviation} \\
        \midrule
        \textbf{H$_4$} & 8 & 4 & $d_1$  & 0.734 & 0.734 & 0.000 \\
        & & & $d_2$  & 0.734 & 0.734 & 0.000 \\
        & & & $d_3$  & 3.000 & \text{$\infty$} & \text{N/A}  \\
        \midrule
        \textbf{H$_2$O$_2$} & 24 & 18 & $d_1$ & 1.029 & 1.000 & 0.029 \\
        & & & $d_2$  & 1.481 & 1.402 & 0.079 \\
        & & & $\theta$ & 0.026 & 0.007 & 0.019 \\
        & & & $\phi$ & 1.418 & 1.400 & 0.018 \\
        \midrule
        \textbf{C$_2$H$_4$O$_3$} & 58 & 20 & $R_y$ & 0.002 & 0.000 & 0.002 \\
        & & & $R_z$  & 0.162 & 0.000 & 0.162 \\

        \bottomrule
    \end{tabularx}
\end{table*}

For the H\(_2\)O\(_2\) molecule, the initial qubit requirement of 24 qubits was reduced to 18 qubits. Four parameters were utilized to describe its geometric configuration, as illustrated in Fig.~\ref{fig:H2O2}(a). These parameters include $d_2$ (the O–O bond length), $d_1$ (the H–O bond length), and $\theta$ and $\phi$ (the azimuthal and polar angles of the H atom, respectively, with the O atom as the origin of the spherical coordinate system). To visualize the optimization trajectory, the parameters $\phi$ and $d_1$ were extracted, as shown in Fig.~\ref{fig:H2O2}(b). The underlying heatmap in Fig.~\ref{fig:H2O2}(b), serving as a standard reference, was generated using classical chemical methods. With initial parameters for $d_1$ and $\phi$ set to 1.0 Å and 0.5 rad, respectively, the trajectory is observed to gradually converge towards the lower energy region of the heatmap, effectively yielding an approximate equilibrium geometry.

\begin{figure*}
\includegraphics[width=\textwidth]{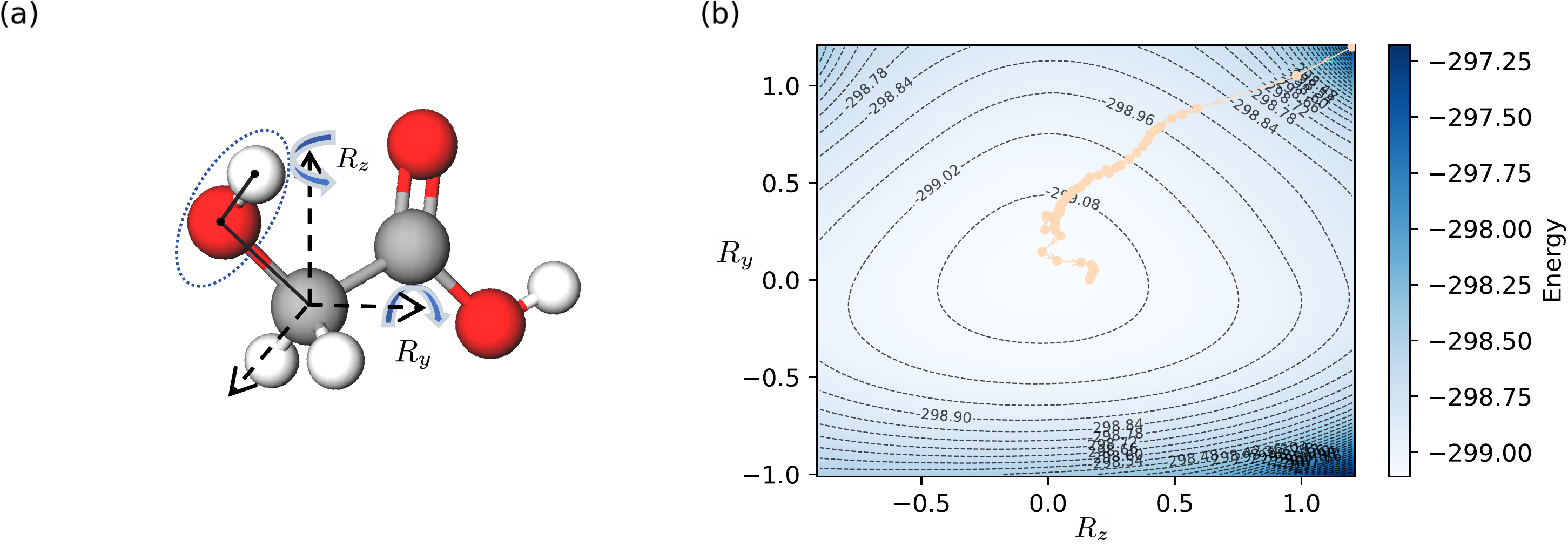}
\caption{Exploration of molecular geometries and configurations using our algorithm. (a) Illustration of the C\(_2\)H\(_4\)O\(_3\) molecule, highlighting the hydroxyl group subject to fixed-point rigid body rotation described by $R_y$ and $R_z$. (b) Trajectory of the C\(_2\)H\(_4\)O\(_3\) optimization in the $R_y$-$R_z$ parameter space, also demonstrating convergence towards an approximate equilibrium geometry.}
\label{fig:C2H4O3}
\end{figure*}

These preceding examples collectively demonstrate the effectiveness of our algorithm in exploring diverse molecular geometries and configurations. Building on this validation, we applied our algorithm to a more complex system: glycolic acid C\(_2\)H\(_4\)O\(_3\), depicted in Fig. 3(a). The dashed blue ellipse in Fig.~\ref{fig:C2H4O3}(b) highlights a hydroxyl (OH) group. The optimization strategy involved the sequential fixed-point rigid body rotation of this OH group around the carbon (C) atom to which it is bonded. The rotation was parameterized by two successive angles: an angle $R_y$ about the y-axis, followed by an angle $R_z$ about the z-axis. Thus, the geometric parameters to be optimized were $R_z$ and $R_y$. Similar to the H\(_2\)O\(_2\) case, Fig.~\ref{fig:C2H4O3}(b) illustrates the optimization trajectory on a heatmap when the initial parameters for $R_z$ and $R_y$ were set to 1.0 rad. The trajectory clearly shows a gradual convergence towards the lower energy region of the heatmap, indicating the successful attainment of an approximate equilibrium geometry for glycolic acid.

\section{Conclusions}

The accurate prediction of equilibrium geometries for large and chemically relevant molecules has long been hindered by the exponential scaling of classical electronic structure methods and the hardware constraints of near-term quantum devices. In this work, we addressed these challenges by introducing a novel hybrid quantum-classical co-optimization framework that tightly integrates Density Matrix Embedding Theory (DMET) with the Variational Quantum Eigensolver (VQE). This approach simultaneously optimizes molecular geometries and quantum circuit parameters, effectively bypassing the costly nested loops of conventional workflows while alleviating qubit requirements.

Our method demonstrated high accuracy and efficiency on benchmark molecules, including H\(_4\)and H\(_2\)O\(_2\), validating its robustness for equilibrium geometry prediction. Crucially, we extended the framework to glycolic acid (C\(_2\)H\(_4\)O\(_3\)), marking the first successful quantum algorithm-based geometry optimization of a molecule of this complexity. This achievement underscores the framework’s capability to dramatically reduce quantum resource demands while preserving chemical accuracy, thereby overcoming two of the most critical bottlenecks in scaling quantum simulations.

By bridging DMET’s fragmentation strategy with VQE’s variational optimization, this work opens a new frontier for quantum chemistry, enabling realistic, large-scale molecular geometry optimization on near-term quantum devices. Beyond its immediate contributions, the framework holds significant promise for application to larger and more diverse chemical systems, including pharmaceutically and industrially relevant molecules. Future research will focus on expanding its scope to periodic materials, incorporating advanced quantum hardware capabilities, and integrating error mitigation and noise-resilient algorithms to further enhance scalability and reliability.

\begin{acknowledgement}

This work is supported by the Innovation Program for Quantum Science and Technology (Grant No.~2023ZD0300200), the National Natural Science Foundation of China NSAF (Grant No.~U2330201) and (Grant Nos. 12361161602 and 92265208), the Innovation Fund of Aerospace Institute 771 (Grant No. 771CX2022003). The scientific calculations  are supported by the High-performance Computing Platform of Peking University.

\end{acknowledgement}

\bibliography{ref}

\end{document}